\documentclass[conference]{IEEEtran}

\usepackage{amsmath}
\usepackage{amssymb}
\usepackage{amsthm}
\usepackage{bm}
\usepackage{algorithm}
\usepackage{mathrsfs}
\usepackage{algpseudocode} 
\usepackage{enumitem}
\usepackage{verbatim}
\usepackage{graphicx}
\usepackage{subfigure}

\newcommand{\citep}[1]{\cite{#1}}

\newtheorem{example}{Example}
\newtheorem{note}{Remark}

\usepackage{algorithm,algcompatible,amsmath}

\usepackage{tikz}
\usepackage{pgfplots}
\usepackage{graphicx}
\usepackage{subfigure}

\pgfplotsset{compat=1.8}

\begin{document}

\title{Secure Regenerating Codes for Reducing Storage and Bootstrap Costs in Sharded Blockchains}

\author{%
	\IEEEauthorblockN{ Divija Swetha Gadiraju, V. Lalitha}
	\IEEEauthorblockA{IIIT Hyderabad \\
		Email: \{divija.swetha@research.iiit.ac.in, lalitha.v@iiit.ac.in\}
		}

	\and
	\IEEEauthorblockN{Vaneet Aggarwal}
	\IEEEauthorblockA{Purdue University \\
		Email: vaneet@purdue.edu
}
}

\maketitle

\begin{abstract}
Blockchain is a distributed ledger with wide applications. Due to the increasing storage requirement for blockchains, the computation can be afforded by only a few miners. Sharding has been proposed to scale blockchains so that storage and transaction efficiency of the blockchain improves at the cost of security guarantee. This paper aims to consider a new protocol, Secure-Repair-Blockchain (SRB), which aims to decrease the storage cost at the miners. In addition, SRB also decreases the bootstrapping cost, which allows for new miners to easily join a sharded blockchain. In order to reduce storage, coding-theoretic techniques are used in SRB. In order to decrease the amount of data that is transferred to the new node joining a shard, the concept of exact repair secure regenerating codes is used. The proposed blockchain protocol achieves lower storage than those that do not use coding, and achieves lower bootstrapping cost as compared to the different baselines.
\end{abstract}

\section{Introduction}

Blockchain technology's unique ability to provide an open ledger for recording transactions while simultaneously ensuring security and verifiability lends itself to a variety of uses.  Blockchain experts envision a huge amount of possible applications, with everything from supply chain management to online personal identification.  Its applications includes  Bitcoin, which is a decentralized cryptocurrency which processes online payments though a peer-to-peer (P2P) system without traversing through a financial institution.  The size of the Bitcoin blockchain has experienced consistently high levels of growth since its creation, reaching approximately 269.82 gigabytes in size as of the end of March 2020 \cite{stat_bc}. The increasing size of blockchain requires high storage capacity at miners which can be afforded only by a few miners \cite{road_scalable_bc}. Further, in order to allow efficient entry for new miners, the amount of data that must be sent to the the new miner (denoted as the bootstrap cost) must be low.  In this paper, we consider coding theoretic techniques to reduce the storage cost as well as the bootstrap cost that would help  more miners to enter the market. 




In order to reduce the  computation burden to solve the proof-of-work (PoW) puzzles, an approach called sharding has been proposed \cite{Elastico}. This paper considers sharded blockchains and apply coding theoretic techniques to reduce the storage cost as well as the bootstrap cost. We present a sharding protocol, called SRB, which is  based on secure regenerating codes that are not only storage efficient but also bandwidth efficient while repairing single node failures. We draw the equivalence between the process of bootstrapping a node and repairing a failed node. Hence, our bootstrap cost is low as compared to  uncoded sharding \cite{rapidchain} and  Secure Fountain Architecture (SeF) based sharding \cite{kadhe2019sef}.

\begin{table*}[ht]
	\centering
	\scalebox{1.2}{
		\begin{tabular}{|c|c|c|c|}
			\hline
			Parameter     &  RapidChain & SeF & SRB \\
			\hline
			Storage Overhead     &   $n_S$     & $(1+\delta)$  & $\frac{n_S \alpha}{L}$ \\
			\hline
			Bootstrap Cost  &   L  & $L + O(\sqrt{L} \log ^2(L/\delta))$ & $\alpha < L$ \\
			\hline
			Security Guarantee & $\frac{n_S}{2}$   &   $n_S - \frac{L + O(\sqrt{L} \log ^2(L/\delta))}{\rho}$   & $\frac{n_S - \alpha}{2}$\\
			\hline
		\end{tabular}
	}
	\vspace*{0.2in}
	\caption{ Performance comparison of RapidChain, SeF based and SRB protocols for sharding.}
	\label{tab:performance}
\end{table*}

In the previous approaches based on codes, the bootstrapping process has two steps: (i) all the blocks are recoverd at the new node (ii) the coded blocks to be stored are again computed as linear combinations of the recovered blocks. In the proposed approach, we bypass the first step by directly trying to recover the coded blocks which are to be stored in the new node. Hence, the bootstrap cost is less as compared to previous method of SeF based sharding.
In the proposed approach, $L$ blocks in the shard are encoded to obtain $n_S\alpha$ blocks and stored on $n_S$  nodes in the shard, with $\alpha < L$ blocks on each node. The code construction uses a product matrix code construction. The bootstrapping problem has connections to the secure repair problem in regenerating codes, and these connections are exploited to provide an efficient bootstrapping transfer.  In addition to the encoding and decoding mechanisms, the entire protocol efficiently integrates the mechanisms of sharding, reference committee reconfiguration and intra-shard consensus. 
The main result can be seen in Table \ref{tab:performance}. As compared to RapidChain, the proposed framework allows for significantly lower storage overhead ($n_S$ vs $\frac{n_S \alpha}{L}$) and bootstrapping cost ($L$ vs $\alpha$). As compared to SeF based sharding, the proposed framework allows for significantly lower bootstrapping cost ($L + O(\sqrt{L} \log ^2(L/\delta))$ vs $\alpha$) with somewhat increase in storage overhead. Further, the performance of the proposed approach for security, encoding complexity  are obtained.



The rest of the paper is organized as follows. Section \ref{sec:related} describes the related work in the areas of sharding and coding-theory based techniques in blockchains. The prior work in the areas that are related to the proposed approach is provided in Section \ref{sec:prelim}.  In Section \ref{sec:sbr}, we will discuss the SBR protocol which combines ideas from rapid chain protocol as well as the secure regenerating codes. We present the performance comparison in Section \ref{sec:performance}. The conclusions are presented in Section \ref{sec:concl}.

\section{Related Work}\label{sec:related}

{\bf Sharding: } Sharding in blockchains was discussed first in Elastico \cite{Elastico}. They proposed to partition the network into shards, each of which processes a disjoint set of transactions. The number of shards grow linearly with the total computational power of the network. Each shard runs a classical byzantine consensus protocol to decide their agreed set of transactions in parallel. In every epoch, a proof-of-work (PoW)  puzzle is solved based on an epoch randomness obtained from the last state of the blockchain. Elastico improved the throughput and latency of Bitcoin. Omniledger \cite{omniledger} is an improvement over Elastico which preserves long term security in blockchains. It introduced atomic cross-shard commit protocol for transactions affecting multiple shards. It ensures security by using a bias-resistant public-randomness protocol.

{\bf Coding Techniques in Blockchains:} Coding theory is applied used to gain storage efficiency in blockchain systems. Recently, rateless codes have been actively applied in various aspects of storage and computing. The authors in \cite{kadhe2019sef}, propose using rateless codes for blockchain storage and for bootstrapping a new node in the network. The fountain codes approach used in  \cite{kadhe2019sef},  has about 1000x storage savings. In \cite{polyshard}, each node stores the coded version of the entire blockchain. So, they achieve up to 30\% storage savings compared to full replication of the blockchain on each node.  In \cite{dynamicStorage_isit}, the authors propose a novel distributed storage code for blockchains as a combination of secret key sharing, private key encryption. Interestingly, they focus on using dynamic zone allocation, which uses a combination of cryptographic and information-theoretic security. The authors in \cite{Erasure_code_low_storagnode}, discuss the possibility of using erasure codes for low storage, especially for light clients. Here, blockchain is  stored as only some coded fragments of each block. In \cite{low_storageRoom}, the authors focus on tackling storage bloating problem with network coding with such a framework does not lose any information. Moreover, with their storage scheme they can save storage room for each node. The problem of designing fault tolerant distributed storage based on blockchains in the context of industrial network environments has been discussed in \cite{liang2020secure}. Here, the authors employ codes with local regeneration in order to repair from single and multiple node failures.

\section{Preliminaries of Sharding and Coded Blockchains}\label{sec:prelim}

\subsection{RapidChain} \label{sec:rapid}

In this section, we will describe the RapidChain protocol which allows scaling of blockchain by sharding. Our approach of using secure regenerating codes in the framework of sharding will be based partly on the ideas discussed here.

The RapidChain protocol for sharding consists of the following elements:

\begin{itemize}
\item {\bf Reference Committee and Epoch Randomness:}
The protocol is divided into epochs.  A reference committee is elected at the start of the protocol.
Reference committee is a set of nodes which are responsible for generating epoch randomness and verifying the PoW puzzles which the new nodes joining the network solve. During the start of each epoch, the reference committee comes to consensus on a reference block which contains the list of all nodes and their assigned shard (We will use the term shard instead of committee in this paper). The reference block is sent to all the nodes.

\item {\bf Committee Reconfiguration and Cuckoo Rule:}
Malicious nodes could strategically leave and rejoin the network and they can take over one of the shard to break the security guarantees of the protocol. Moreover, we assume a byzantine adversary that can actively corrupt a constant number of honest nodes. In order to prevent this attack, shards have to be periodically reconfigured faster than the adversaryÕs ability to generate churn. The communication overhead is very high if all the committees need to re-elected.
RapidChain follows the Cuckoo rule to shuffle only a subset of nodes during the reconfiguration at the beginning of each epoch. RapidChain ensures that shards are balanced with respect to their sizes as nodes join or leave the network. To map the nodes to shards, each node is mapped to a random position in $(0,1]$. Then, the range is partitioned into $m$ regions. In the Cuckoo rule, when a node wants to join the network, it is placed at a random position, while all nodes in a constant-sized interval surrounding the new node's position are moved to new random positions in the interval $(0,1]$.

\item {\bf Ledger Pruning:} Ideally, a new node joining the committee has to download all the blocks in the shard. However, since the throughput of the system is high, the amount of download is quite high. To avoid this, the nodes which join the network prune the ledger and store. The disadvantage of pruning is that there are only a few archival nodes which need to be contacted if the complete ledger needs to be downloaded.

\item {\bf Intra-committee Consensus and Cross-shard Transactions:} The intra-committee consensus is accomplished by a byzantine consensus protocol based on fast gossip algorithm with application of erasure codes. Cross-shard transactions are performed based on a certain inter-committee routing scheme.

\end{itemize}



\subsection{Secure Fountain Codes for Sharded Blockchains} \label{sec:sef}

RapidChain has high storage overhead since every node in the shard stores a full copy of the blocks in the shard. To save on the storage costs, one of the existing solutions is based on Secure Fountain codes (SeF). 
The authors of \cite{kadhe2019sef}  proposed an architecture, called Secure Fountain (SeF), which is  based on fountain codes and
enables any full node to encode validated blocks into a small number of coded blocks, thereby reducing its storage costs by orders of magnitude. One of the key idea in SeF is to
use the header-chain as a side-information to check whether a coded block is maliciously formed while it is getting decoded. The coded blocks in SeF are called droplets, the nodes storing coded blocks are called droplet nodes, and any new node joining the system is called a bucket node. During bootstrap, a  bucket
node collects sufficiently many droplets and recovers the blockchain even when some droplet nodes are
adversarial, providing murky (malicious) droplets. After validating the blockchain, a bucket node will
perform encoding to turn itself into a droplet node. In this way, droplet nodes will slowly replace archival
nodes.

{\bf Encoding:} The encoding is performed  using a Luby Transform (LT) code \cite{luby2002lt}.  LT codes admit a computationally efficient decoding procedure known as peeling decoder (also
known as a belief propagation). SeF exploits 
peeling process  to introduce resiliency against maliciously formed blocks by using the
header-chain as a side-information and leveraging Merkle roots stored in block-headers, which is explained as follows.

{\bf Peeling Decoder:} Consider a bucket node that is interested in recovering the blockchain $B$. It contacts an arbitrary subset
of $n$ ($n \ge k$) droplet nodes, and downloads the stored data. This includes droplets $C_j$ and vectors $v_j$.
Without loss of generality, let us (arbitrarily) label the downloaded droplets as $C_1, C_2, \cdots, C_{ns}$. Note that,
since a coded droplet does not have any semantic meaning, the bucket node cannot differentiate between
the honest and malicious droplets within the downloaded ones. We assume that the bucket node has access to the honest header-chain. Note that this can be
done by contacting several droplet nodes, and obtaining the longest valid header-chain as described in \cite{kadhe2019sef}. The decoding proceeds in iterations. In each iteration the algorithm decodes (at most) one block until
all the blocks are decoded, otherwise the decoder declares failure. Note that Step (3) differentiates the error-resilient peeling decoder from the classical peeling decoder for
an LT code. More specifically, the classical peeling decoder always accepts a singleton, whereas
the error-resilient peeling decoder may reject a singleton if its header and/or Merkle root does not match
with the one stored in the header-chain.

 SeF codes allow the network to tune the storage savings as a parameter, depending upon
how much bootstrap cost new nodes can tolerate. When SeF codes are tuned to achieve $k$-fold storage
savings, a new node is guaranteed to recover the blockchain with probability $(1 - \delta)$ by contacting
$k +O(\sqrt{k} ln^2(k/\delta))$ honest nodes.

\section{Secure Repair Block in Sharded Blockchains} \label{sec:sbr}

In our work, we assign nodes to shards and perform our coding scheme on individual shards. We use the procedure followed in  \cite{rapidchain} to execute the random assignment of nodes to shards. In the following, we describe the components of the Secure Repair Block protocol and also algorithms corresponding to encoding, bootstrapping and reconstruction.

 \begin{figure}
     \centering
     \includegraphics[width=3in]{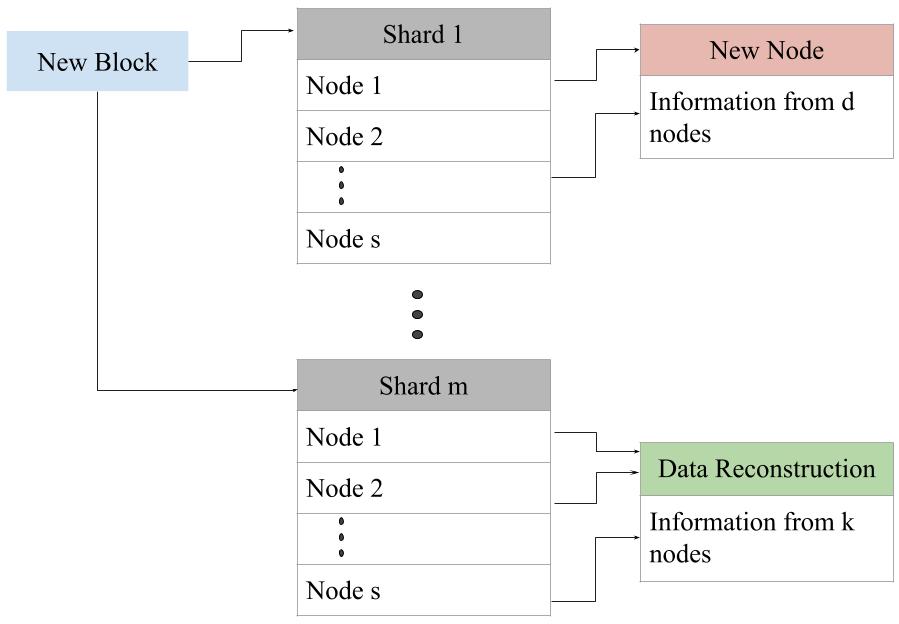}
     \caption{Illustration of Secure Repair Block protocol in blockchains.}
     \label{fig:sharding}
 \end{figure}
 
 \subsection{Components of Secure Repair Block Protocol}
 
We build the secure repair block protocol around the RapidChain protocol. We have the following components:
 \begin{itemize}
 \item Reference committee is elected at the start of the protocol, which help in generation of epoch randomness and creating reference block. The reference block in addition to the epoch randomness and a list of nodes, has additional information of ``Node Encoder Coefficient". The node encoder coefficient will be described later.
 \item Committee reconfiguration is performed using Cuckoo rule described before.
 \item Intra-shard consensus and cross-shard transactions are performed similar to RapidChain.
 \item The blocks are encoded in a certain way and multiple coded blocks ($\alpha$ of them) are stored in a single node. If a new node joins the system, bootstrapping the node is considering equivalent to repairing the node as shown in Fig. \ref{fig:sharding}. The code construction which will facilitate this process and will be described in the rest of the section.
 \item It has been claimed in \cite{avarikioti2019divide} that to achieve scalability, any robust sharded ledger has to perform compaction of state. Hence to perform verification, we propose that a node contacts other nodes to download a certain number of coded blocks and recovers the original blocks. 
 \end{itemize}
  
 We would like to note that we do not consider ledger pruning as part of Secure Repair Block protocol.


\subsection{Secure MBR Regenerating Codes} \label{subsec:sec_mbr}
\begin{figure}
     \centering
     \includegraphics[width=3.2in]{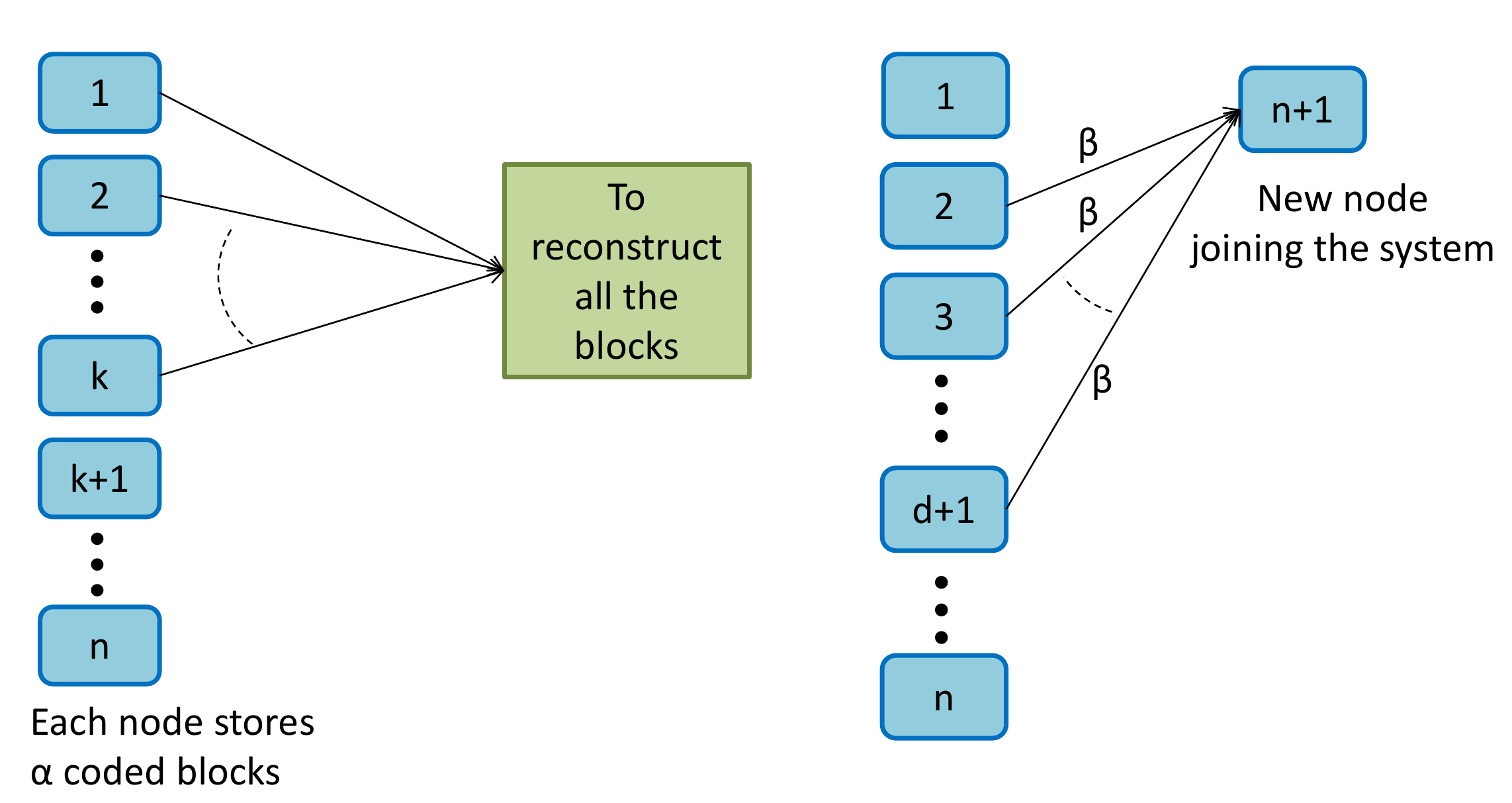}
     \caption{Regenerating code framework in the context of blockchains.}
     \label{fig:regen_code}
 \end{figure}

In this section, we describe the framework of secure regenerating codes which guarantee minimum bandwidth during the repair of a single node. These class of codes are known as minimum bandwidth regeneration (MBR) codes.
The code construction as well as the security properties of the code have been derived in \cite{infTh_secure_regenerating}. These are discussed in detail here for the sake of completeness.
These codes will be used in a later section to provide a storage scheme for blockchains with sharding.
Consider a file of size $L$ over a finite field $\mathbb{F}_q$ with the corresponding message symbols denoted by $m_1, m_2, \ldots, m_L$. These $L$ symbols are encoded into $n \alpha$ symbols and stored at $n$ nodes each of which can store $\alpha$ symbols over a finite field $\mathbb{F}_q$. We need to be able to perform two functions on this kind of a distributed storage system.
\begin{itemize}
\item {\bf Data Reconstruction:} By connecting to any $k$ nodes and downloading all the $\alpha$ symbols present in the $k$ nodes, we have to be able to recover all the $L$ message symbols.
\item {\bf Node Repair:} By connecting to any $d$ nodes and downloading $\beta < \alpha$ symbols from the $d$ nodes, we would like to recover the contents of a single node.
\end{itemize}
The framework of regenerating code in the context of blockchains is shown in Fig. \ref{fig:regen_code}.

We consider the threat
model wherein the contents of one or more nodes may be malicious.  A node that is
malicious may send arbitrarily corrupted data during data reconstruction and data repair.

For the case of minimum bandwidth regenerating codes $\alpha = d\beta$. In the current code construction, we consider $\beta =1$ and hence $d = \alpha$.
The code construction is given by the following product-matrix framework:
\begin{equation}
C = \Psi M,
\end{equation}
where $\Psi$ is encoding matrix of size $n \times \alpha$ and $M$ is message matrix of size $\alpha \times \alpha$. 
Each row of $C$ is a set of $\alpha$ symbols which is stored in a node. 
The message matrix $M$ has the following symmetric structure:
\begin{equation} \label{eq:msg_matrix}
M = [m_{i,j}] = \begin{bmatrix} U & V \\  V^T & 0 \end{bmatrix},
\end{equation}
where $U$ is a symmetric matrix of size $k \times k$ and $V$ is a matrix of size $k \times (\alpha-k)$. The number of variables in the above matrix is thus given by $L = k(\frac{k+1}{2}) + k (\alpha - k) $. One possible choice of the encoding matrix is that of Vandermonde matrix given below. 
\begin{equation}
\Psi = \begin{bmatrix} 1 & \gamma_1 & \gamma_1^2 & \ldots & \gamma_1^{\alpha-1} \\
1 & \gamma_2 & \gamma_2^2 & \ldots & \gamma_2^{\alpha-1} \\
\vdots & \vdots & \ddots & & \vdots \\
1 & \gamma_n & \gamma_n^2 & \ldots & \gamma_n^{\alpha-1}
\end{bmatrix},
\end{equation}
where all $\gamma_i$ are distinct elements of the finite field $\mathbb{F}_q$.
With the encoding matrix chosen as Vandermonde matrix, the above code construction can also be interpreted as being obtained by evaluating polynomials at various points in the finite field $\mathbb{F}_q$.
Consider the following set of $\alpha$ polynomials, 
\begin{equation} \label{eq:msg_poly}
P^{(i)}(x) = \sum_{j=0}^{\alpha-1} m_{i,j} x^j.
\end{equation}
It can seen that the code can be obtained by evaluating the above polynomials at distinct elements $\gamma_1, \gamma_2 \ldots, \gamma_n$ of the finite field $\mathbb{F}_q$, $q \geq n$. We would like to note that this code construction is very flexible in terms of addition of nodes as long as $k$ and $\alpha$ are kept constant. This is because we just need to add another evaluation point $\gamma_{n+1}$ to the existing code and the resultant code is still an MBR code comprising $n+1$ nodes. This property will be used later in the case of blockchains to bootstrap new nodes.


\begin{note}
There are also other class of codes known as minimum storage regenerating codes (MSR codes) which can be used to reduce the bootstrap cost as well. These codes offer a different operating point with respect to storage cost and bootstrap cost. These codes have lesser storage cost than MBR codes with an increase in the bootstrap cost. However, the code construction for a range of parameters is also based on product matrix framework and is similar to the MBR code. Thus, we will not discuss this variant of regenerating codes in this paper.
\end{note}

\subsection{Encoding}
We analyze the system based on rounds or epochs.  At the beginning of each epoch, a reference committee agrees on a reference block consisting of the list of all active nodes for that epoch as well as their assigned shards. 
In addition to assigning nodes to the shards, in our protocol, the reference committee also assigns $n_S$ distinct coefficients ($\{ \gamma_i \}$ termed as node encoder coefficients) from the finite field to every node in the shard. This information is also stored in the reference block.
During the epoch, $L = k\alpha - {\binom k 2}$ blocks in the shard are encoded to obtain $n_S \alpha$ blocks and stored on $n_S$ nodes in the shard. The encoding process is described in Algorithm \ref{alg:encoding}.

\begin{algorithm}
	\caption{Coding the blocks in a shard}\label{alg:encoding}
	\begin{algorithmic}[1]
		\STATE Input $B_1, \ldots, B_L$ which are $L$ input blocks
		\STATE Stripe the blocks into units of the finite field $\mathbb{F}_q$. The stripes of block $B_{\ell}$ are denoted by $B_{\ell,1}, \ldots, B_{\ell,Z}$. $Z$ is number of stripes in a block, which can determined by the block size and the field size.
		\STATE Form $\alpha$ message polynomials as described in \eqref{eq:msg_poly} with $\{B_{\ell}\}$ in place of $\{m_{i,j}\}$.
		\STATE Evaluate the $\alpha$ message polynomials at the coefficient $\gamma_i$ assigned by the reference committee to obtain the coded stripes $C_{i,j,s}, 1 \leq j\leq \alpha, 1 \leq s \leq Z$.
		\STATE Merge the stripes corresponding to a single code block and form $\alpha$ code blocks denoted by $C_{i,1}, C_{i,2}, \ldots, C_{i,\alpha}$.
		\STATE The coded blocks $C_{i,1}, C_{i,2}, \ldots, C_{i,\alpha}$ are stored along with coefficient $\gamma_i$.
	\end{algorithmic}
\end{algorithm}

To understand the encoding process as the blocks are coming in, the first round of encoding is done when the number of blocks in the epoch is $L$. Then the first $L$ blocks are encoded using the procedure described above and the original blocks are deleted. The second round of encoding is performed when the number of blocks in the epoch is $2L$. In the second round, blocks from $L+1, \ldots, 2L$ are encoded using the procedure described above and the process continues.

\subsection{Boostrapping new nodes to a shard}
The code described in Algorithm \ref{alg:encoding} has the data reconstruction and node repair properties which have been listed in the specifications of a regenerating code. Now, we will perform secure node repair with the same code by contacting more number of helper nodes.
Here,  we describe a procedure for repairing a single node failure in the presence of $p$ malicious nodes. As can be seen for the code construction, a node $i$ can be identified by its coefficient $\gamma_i$. In order to repair a node whose coefficient is $\gamma_i$, $d+2p$ other nodes whose coefficients are say $1, 2, \ldots, {d+2p}$ (all different from $i$) send the following symbols:
\begin{equation}
[1 \ \  \gamma_i \ \ \gamma_i^2 \ldots \ \ \gamma_i^{\alpha-1}]M[1 \ \ \gamma_j \ \ \gamma_j^2 \ldots \ \  \gamma_j^{\alpha-1}]^T,
\end{equation}
where $j = 1,2,\ldots, d+2p$. By collecting the $d+2p$ symbols, the resultant vector is equivalent to encoding the vector $[1 \ \  \gamma_i \ \ \gamma_i^2 \ldots \ \ \gamma_i^{\alpha-1}]M$ with an MDS code of length $d+2p$ and dimension $d = \alpha$. Hence, it can result of correct output of $[1 \ \  \gamma_i \ \ \gamma_i^2 \ldots \ \ \gamma_i^{\alpha-1}]M$ even if $p$ nodes are malicious and send erroneous symbols.

When a new node joins a shard, the contents of the node are obtained by repair process. This is accomplished by the new node contacting $d+2p$ nodes and downloading one coded block from each one of them. The reference committee keeps track of which nodes have which evaluation points and assigns a unique coefficient $\gamma_i$ to the new node. In the process of reconfiguration of shards/bootstrapping, we are assuming that at any point there are at least $d+2p$ nodes in any shard. Also, this scheme is flexible in terms of the number of malicious nodes we can tolerate. If an estimate of level of security is known at any point of time, then we can contact lesser number of nodes (since $p$ is less) and perform the bootstrapping operation.
 The data sent by the peers corresponds to the result of one linear combination of the $\alpha$ coded blocks contained in the stored data. Once the new node receives $(d+2p)$ coded blocks, it applies a decoding procedure to recover its own coded blocks. The algorithm to bootstrap a new node to shard is described in Algorithm \ref{alg:repair}.

\begin{algorithm}
	\caption{Bootstrapping new node to a shard}\label{alg:repair}
	\begin{algorithmic}[1]
		\STATE Input coefficient $\gamma_i$ which corresponding to the evaluation point of the new node $i$.
		\STATE The new node $i$ queries $d+2p$ other nodes in the shard with its coefficient $\gamma_i$.
		\STATE $j^{\text{th}}$ node, $ 1 \leq j \leq d+2p$ sends the following linear combination to the new node $[1 \ \  \gamma_i \ \ \gamma_i^2 \ldots \ \ \gamma_i^{\alpha-1}]M[1 \ \ \gamma_j \ \ \gamma_j^2 \ldots \ \  \gamma_j^{\alpha-1}]^T$ after the message matrix $M$ is replaced with the corresponding blocks and the process of striping similar to that in the encoding algorithm. 
		\STATE After having received $d+2p$ coded blocks from the $d+2p$ nodes in the shard, the new node performs a $(d+2p, d)$ Reed Solomon decoding to obtain the coded blocks which it has to store.
	\end{algorithmic}
\end{algorithm}

\begin{figure}
     \centering
     \includegraphics[width=3.2in]{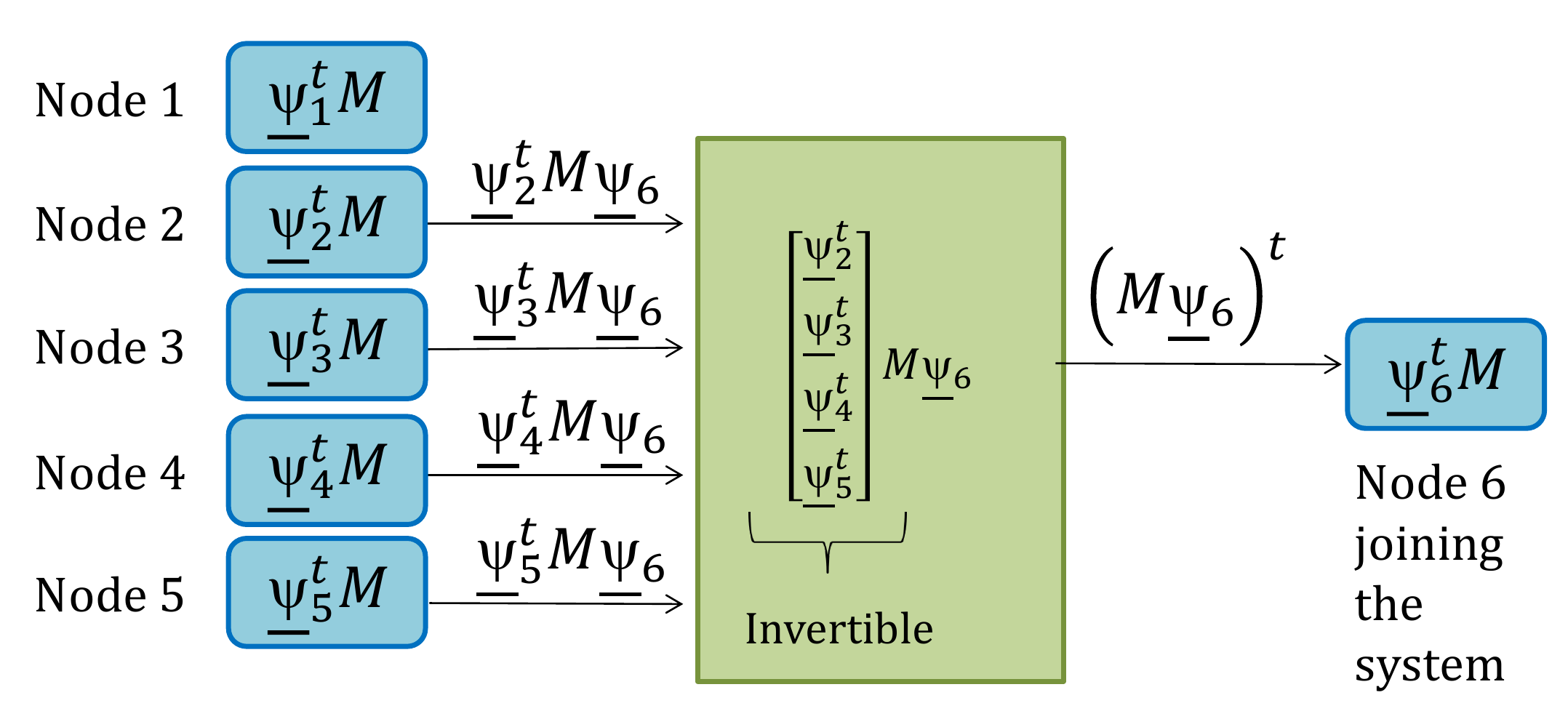}
     \caption{Bootstrapping a node is equivalent to repair of product-matrix MBR code.}
     \label{fig:mbr_repair}
 \end{figure}

\begin{example}
We illustrating the process of bootstrapping a node (in the absence of adversarial nodes) through an example. Let there be 5 nodes in the system such that $k=3$ and $d = \alpha = 4$. $M$ is a message matrix consisting of $L = k\alpha - {\binom k 2} = 9$ blocks arranged as follows:
\begin{equation}
M = \begin{bmatrix} B_1 & B_2 & B_3 & B_7 \\  B_2 & B_4 & B_5 & B_8 \\  B_3 & B_5 & B_6 & B_9 \\  B_7 & B_8 & B_9 & 0 \end{bmatrix}
\end{equation}
The $i^{\text{th}}$ node stores $4$ coded blocks $[1\ \  \gamma_i \ \ \gamma_i^2 \ \ \gamma_i^3]M$. Suppose if a $6^{\text{th}}$ node joins the system, then in order to compute $[1 \ \ \gamma_6 \ \ \gamma_6^2 \ \ \gamma_6^3]M$, the new node downloads one block each from $4$ nodes. For example, in Fig. \ref{fig:mbr_repair}, nodes $2,3,4,5$ send one block each given by $[1 \ \ \gamma_i \ \ \gamma_i^2 \ \ \gamma_i^3]M\begin{bmatrix} 1 \\ \gamma_6 \\ \gamma_6^2 \\ \gamma_6^3 \end{bmatrix}$ and then the new node computes $[1 \ \ \gamma_6 \ \ \gamma_6^2 \ \ \gamma_6^3]M$ from these $4$ blocks.
\end{example}

\subsection{Reconstruction}

Here,  we describe a procedure for reconstructing data in the presence of $p$ malicious nodes. For data reconstruction, we will contact $k+2p$ nodes and download all $\alpha$ symbols from each of them. Thus, we have access to the following symbols:
\begin{eqnarray}
\Psi_{dr} M & = & [\Delta_{dr} \ \Phi_{dr}] \begin{bmatrix} U & V \\  V^T & 0 \end{bmatrix} \\
& = & [\Delta_{dr} U + \Phi_{dr} V^T \ \ \Phi_{dr} V],
\end{eqnarray}
Where $\Psi_{dr}$ refers to an arbitrary $(k+2p) \times \alpha$ submatrix of $\Psi$. The reconstruction of the message matrix involves two steps:
\begin{itemize}
\item In the first step, $ \Phi_{dr} V$ can be used to be recover $V$ in the presence of $p$ mailcious nodes. This is because $ \Phi_{dr} v_j$ for every column vector $v_j$ in $V$ forms an MDS code with parameters $(k+2p, k)$. Hence, $p$ errors can be tolerated. By applying similar procedure to every column of $V$, we can recover $V$ completely.
\item In the second step, the contribution due to $V$ is subtracted from the term $\Delta_{dr} U + \Phi_{dr} V^T$ to result in $\Delta_{dr} U$. We can recover $U$ using the property of MDS code similar to that in the above step.
\end{itemize}


\section{Performance Comparison} \label{sec:performance}

In this section, we will compare the performance of three schemes: RapidChain, SeF based sharding, and SRB protocol. The performance metrics include (i) storage overhead, (ii) bootstrap cost, (iii) encoding complexity, and (iv) security guarantee.

\subsection{Storage Overhead}

In the case of RapidChain, since each block in the shard is replicated $n_S$ times where $n_S$ is the number of nodes in the shard, the storage overhead is $n_S$. In the case of SeF based sharding, the storage overhead is $(1+\delta)$ where $\delta$ is some small positive quantity. Here, we would like to note that in SeF based sharding, we need to combine a large number of blocks (order of thousands) in order to obtain a code which can be decoded successfully, i.e., $L$ has to be large. In the case of SRB protocol, the storage overhead is given by $\frac{n_S \alpha}{L}$, which is lower than that of RapidChain and higher than that of SeF-based sharding since $\alpha < L$.

\subsection{Bootstrap Cost}

In the case of RapidChain, a new node joining the network has to download the entire blockchain in that shard. So, the bootstrap cost will be $L$ blocks. In the case of SeF based sharding, a new node joining the network has to download at least $L + O(\sqrt{L} \log ^2(L/\delta))$ blocks. Bootstrap cost for SeF based sharding is high because the new nodes first recovers all the $L$ blocks and then later computes the linear combination which it has to store. In secure regenerating code based sharding protocol, we have to download $\alpha$ blocks, where $\alpha < L$. We get this advantage since we are considering the bootstrap of a new node as a single node repair problem and we are downloading just as much content needed to recover the coded blocks to be stored in that node.

\begin{example}
Consider a case when the block size is 2MB. There are 16000 nodes in the system with 16 shards and hence 1000 nodes per shard. We assume that $L = 1065$ blocks have been processed per shard.The storage per node for RapidChain (without ledger pruning) is 2.13GB and bootstrap cost is also 2.13GB. The storage per node for SeF based sharding is 4MB (Assuming number of coded blocks stored per node $\rho = 2$) and the bootstrap cost is $>2.13GB$. The parameters for the SRB protocol are given by $k=30$, $\alpha = 50$, $L = k\alpha - {\binom k 2} = 1065$. With these parameters, the storage per node is 100MB and the bootstrap cost is also 100MB. Hence, the SRB protocol saves both in terms of storage and bootstrap cost.
\end{example}

\subsection{Epoch Security}

Let $t_S$ be the maximum number of malicious nodes within a shard which can be tolerated by the protocol.  
The number of malicious nodes which can be tolerated in the case of RapidChain is $t_S=\frac{n_S}{2}$ where $n_S$ is the number of nodes within a shard. 
For the SeF based sharding and SRB protocol, security guarantee is given by the maximum number of malicious nodes in the presence of which a new node joining the shard is bootstrapped with correct coded blocks.  
In a SeF based sharding scheme, the number of malicious nodes which can be tolerated are $t_S = n_S - \frac{L + O(\sqrt{L} \log ^2(L/\delta))}{\rho}$. For the SRB protocol, the number of malicious nodes which can tolerated are $t_S = \frac{n_S - \alpha}{2}$.

Consider a blockchain network with $N$ nodes and $m$ shards. Each shard has $n_S$ nodes in it. Let $T$ be the total number of malicious nodes in the network. A random variable $X$ represents the number of malicious nodes. The probability that the committee election scheme fails in the first epoch be $p_{bootstrap}$. The value of $p_{bootstrap} \leq 2^{-26.36}$ in \cite{rapidchain}. In \cite{MAth_securityModel_new}, the security analysis of sharding schemes is analyzed assuming a hypergeometric distribution for assigning nodes to the shards. Cumulative hypergeometric distribution, $H(N,P,n_S,p_S)$, is used in calculating the failure probability of one shard and also for failure probability of a given sharding scheme.
\begin{equation}
    H(N,T,n_S,t_S)= \sum_{l=t_S }^{n_S} \frac{\binom T l \binom  {N-T} {n_S-l}}{\binom N {n_S}}
\end{equation}
Hoeffding bound is proved to provide a tight bound for failure probability in \cite{MAth_securityModel_new}, which is given by
\begin{equation}
     H(N,T,n_S,t_S)  \leq G(x),
\end{equation}
where $G(x) = \left(\left( \frac{g}{r} \right)^r\left ( \frac{1-g}{1-r}\right )^{1-r}\right)^{n_S}$, where $r= \frac{t_S}{n_S}$.
Let $p_s$ be the probability of a shard failure, which means number of malicious nodes exceed the limit. The failure probability of Rapidchain \cite{MAth_securityModel_new} is upper bounded by 
\begin{equation}
    p_{bootstrap} + m p_s \leq U(x), \label{upperbund}
\end{equation}
where $U(x) = p_{bootstrap} + m G(x)$.

The performance metrics of storage overhead, bootstrap cost and epoch security have been summarized in Table \ref{tab:performance}.

\subsection{Encoding Complexity}

In RapidChain, there is minimal encoding complexity which involves only copying of blocks from one node to the other. In SeF based sharding, the encoding in every node involves just binary field operations, whose complexity is again very less. In the secure regenerating code based sharding, encoding in the initialization phase involves computing one row of the matrix product which involves taking $\alpha^3$ multiplications over the finite field. In the reconfiguration/bootstrapping phase, computing the contents of a node involves decoding a $(r = d+2p, d)$ Reed Solomon code whose complexity is given by $r^2\ log^2 r\ log logr$.

\subsection{Latency and Throughput}

Latency of the SRB protocol will be approximately same as that of RapidChain. The reason is two-fold. The consensus protocol is synchronous and happens in 4 rounds for both the protocols. The additional delay incurred by the SRB protocol is due to contacting a set of nodes and downloading coded blocks in order to recover original blocks. This delay is small compared to the $\delta$ set for the synchronous protocol and hence the latency for reaching consensus would be nearly same as that of the RapidChain protocol. Thus, we have that $\tau_{SRB} \approx \tau_{RC}$. The throughput factor for the RapidChain in terms of the latency and the number of shards is given in \cite{avarikioti2019divide}. The throughput factor for the SRB protocol can be expressed in terms of its latency exactly using the same expression and is given by
\begin{equation}
\sigma_{SRB} < \mu  \ \tau_{SRB} \left ( \frac{n}{\ln n} \right ) \left ( \frac{(a_{SRB} - p)^2}{2 + a_{SRB} - p} \right ) \left ( \frac{1}{v} \right ),
\end{equation}
where
\begin{itemize}
\item $\mu$ is the ratio of honest blocks in the chain to the total number of blocks, 
\item $n$ is the total number of nodes and $\ln n $ is the number of nodes in a shard
\item $a_{SRB} = \frac{1}{2} - \frac{\alpha}{2 \ln n}$ is the resiliency of the SRB protocol within a shard
\item $p$ is the fraction of malicious nodes in the system
\item $v$ is the average size of transactions
\end{itemize}
We can see that the throughput factor of SRB is slightly less than that of RapidChain because $a_{SRB} < \frac{1}{2}=a_{RC}$.

\subsection{Other Aspects}

Here, we would like to point out two differences between SeF based sharding and secure regenerating codes based sharding. The encoding process for SeF based sharding is probabilistic in nature and hence there is a chance (though extremely less) that the encoding is such that we cannot recover the original blocks from the encoded blocks. On the contrary, secure regenerating codes based sharding protocol is deterministic and hence the blockchain data is definitely present in the system. 

Secondly, SeF based sharding is completely decentralized in nature. Secure regenerating code based sharding is not completely decentralized in the sense that a reference committee has to keep track of the assignment of node encoder coefficients to nodes (negligible overhead is incurred in this process) and if a new node joins the shard, it has to be assigned a coefficient which is not previously assigned to any node in the shard.

Finally, in SeF based sharding, honest header chain is required as side information for applying the peeling based decoder. In secure regenerating code based sharding, there is no need to have the honest header chain as side information as the decoding process is via error correction capability of the code.

\section{Conclusions and Future Work}\label{sec:concl}

In this paper, we have presented a secure-repair-block protocol based on regenerating codes to reduce both storage and bootstrap costs. The proposed protocol is based on a sharding approach which uses efficient secure regenerating codes that are not only storage efficient but also bandwidth efficient while repairing failures. Drawing connections between regenerating codes and blockchain, significant storage savings and savings in bootstrapping costs are obtained, which makes it easier for a new miner to enter the system (since storage is one of the key challenge that requires only larger capability miners to enter). 

This paper demonstrates improvements in storage that allow for easy entry of the lower-capability miners, a detailed implementation of SRB protocol is left for the future. Such implementation can help see the latency and throughput performance of the system.
\section*{Acknowledgement}

Divija Swetha Gadiraju would like to acknowledge the support of Qualcomm
Innovation Fellowship, India.

\bibliographystyle{IEEEtran}


\bibliography{refs}

\end{document}